\def\bc{\begin{center}}                \def\ec{\end{center}}
\def\be{\begin{equation}}              \def\ee{\end{equation}}
\def\bear{\begin{eqnarray}}            \def\eear{\end{eqnarray}}
\def\la{\langle}      \def\ra{\rangle}     \def\l{\left}
\def\r{\right}        \def\dg{\dagger}      \def\ci{\cite}
\def\alf{\alpha}      \def\lb{\label}     
         \def\bet{\beta}
\def\dlt{\delta}         \def\tld{\tilde}

  \def\vs{\vspace}   \def\pr{\prime}
\def\bt{\begin{tabular}}          \def\et{\end{tabular}}

\documentstyle[12pt]{article}
\begin{document}

\begin{flushright}                    e-print quant-ph/9706028\\
                            Preprint INRNE-TH-97/5 (12 June 1997)
\end{flushright}
\bigskip
\begin{center}
    {\Large \bf Barut-Girardello coherent states for $sp(N,C)$\\
and multimode Schr\"odinger cat states } \end{center}
\medskip
\centerline{ D.~A. Trifonov\footnote[0]{$^+$e-mail:
3fonov@phys.acad.bg}$^{+}$}
\centerline{Institute of Nuclear Research,}
\centerline{72 Tzarigradsko Chaussee,\,\,1784 Sofia, Bulgaria}
\medskip \bigskip

\begin{center}
\begin{minipage}{14cm}
\centerline{\small {\bf Abstract}}    \medskip \bigskip

{\small
Overcomplete families of states of the type of Barut-Girardello coherent
states (BG CS) are constructed for noncompact algebras $u(p,q)$ and
$sp(N,C)$ in quadratic bosonic representation. The $sp(N,C)$ BG CS  are
obtained in the form of multimode ordinary Schr\"odinger cat states.  A
set of such macroscopic superpositions is pointed out which is
overcomplete in the whole $N$ mode Hilbert space (while the associated
$sp(N,C)$ representation is reducible). The multimode squared amplitude
Schr\"odinger cat states are introduced as macroscopic superpositions of
the obtained $sp(N,C)$ BG CS.} \end{minipage} \end{center} \vspace{1cm}

%%Section 1
\section{Introduction}

Recently an interest is shown in the literature
\ci{Trif94,Vourdas,Trif96,Brif97,Trif97,Fujii} to applications and
generalizations of the Barut-Girardello coherent states (BG CS) \ci{BG}.
The BG CS were constructed as eigenstates of lowering Weyl operator of the
algebra $su(1,1)$. The BG CS representation was used to construct
explicitly squeezed states (SS) for the generators of the group $SU(1,1)$
which minimize the Schr\"odinger uncertainty relation for two observables
\ci{Trif94} and eigenstates of general element of the complexified 
$su(1,1)$
\ci{Trif96,Brif97}.  These algebra related CS can be considered
as states which generalize the eigenvalue property of BG CS to the case of
linear combination of lowering and raising Weyl operators and even of all
the $SU(1,1)$ generators.  Passing to other algebras it is important first
to construct the eigenstates of Weyl lowering operators, which is the
extension of the BG definition of CS to the desired algebra.

Our aim in the present work is to construct BG CS for the symplectic
algebra $sp(N,C)$ and its subalgebras $u(p,q),\,\,p+q=N$,  in the
quadratic bosonic representation. Here $N$ is the dimension of Cartan
subalgebra, while the dimension of $sp(N,C)$ is $N(2N+1)$, $N=1,2,\ldots$,
\ci{BaRo}.  We establish that the $sp(N,C)$ BG CS take the form of
superpositions of multimode Glauber CS \ci{Glauber} $|\vec{\alf}\ra$ and
$|-\vec{\alf}\ra$ (Eq.  (\ref{sp(N,C)bgcssolu})), i.e. the form of
multimode ordinary Schr\"odinger cat states.  The set of these macroscopic
superpositions of Glauber (or canonical \ci{KlSk}) CS includes several
subsets of states extensively studied in quantum optics (see e.g.
\ci{cats1,cats2}).  We also introduce {\it multimode squared amplitude
Schr\"odinger cat states} as superpositions of the constructed $sp(N,C)$
BG CS.

In the recent E-print \ci{Fujii} the BG CS have been constructed
for the $u(N-1,1)$ algebra.  Here we construct overcomplete families of
states for $u(p,q)$.

%%% Section 2
\section{The Barut-Girardello coherent states}

The property of the Glauber CS $|\alf\ra$ to be eigenstates of photon
number lowering operator $a$, $a|\alf\ra = \alf|\alf\ra$ ($\alf$ is
complex number, $[a,a^\dg] = 1$) was extended by Barut and Girardello
\ci{BG} to the case of Weyl lowering operator $K_-$ of $su(1,1)$ algebra.
Here we briefly review some of their properties. The defining equation is

  \be\lb{bgcsdef}
  K_-|z;k\ra = z|z;k\ra,
  \ee
where $z$ is (complex) eigenvalue and $k$ is Bargman index. For discrete
series $D^{(\mp)}(k)$ $k$ takes the values $\pm1/2,\,\pm1,\,\ldots$. The
Cartan-Weyl basis operators $K_{\pm}=K_1\pm iK_2,\,\, K_3$ of $su(1,1)$
obey the relations

  \be\lb{su11comrel}
  [K_3,K_\pm] = \pm K_\pm, \,\,\,\, [K_-,K_+] = 2K_3,
  \ee
with the Casimir operator $C_2 = K_3{}^2 -(1/2)[K_-K_+ + K_+K_-] =
k(k-1)$.  The expansion of these states over the orthonormal basis of
eigenstates $|k+n,k\ra$ of $K_3$ ($K_3|n+k,k\ra = (n+k)|n+k,k\ra$, $n =
0,1,2,\ldots$) is

\be\lb{bgcsexpan}
|z;k\rangle =
\frac{z^{k-1/2}}{\sqrt{I_{2k-1}(2|z|)}} \sum_{n = 0}^{\infty}
\frac{z^{n}}{\sqrt{n!\Gamma(2k+n)}} |n+k,k\rangle ,
\ee
where $I_\nu(z)$ is the first kind modified Bessel function,  and
$\Gamma(z)$ is gamma function.  The above BG states are normalized to
unity. Their scalar product is

$$\la k;z_{1}|z_{2};k \ra =
I_{2k-1}(2\sqrt{z_{1}^{\ast}z_{2}})\l[\sqrt{ I_{2k-1}(2|z_{1}|)
I_{2k-1}(2|z_{2}|) }\r]^{-1}$$
and they resolve the identity operator,

\be\lb{bgcsresolu}
\int d\mu(z,k) |z;k
  \ra\la k;z| = I, \qquad d\mu(z,k) =
  \frac{2}{\pi} K_{2k-1}(2|z|) I_{2k-1}(2|z|) d^{2}\! z ,
  \ee
where $K_{\nu}(x)$ is the modified Bessel function of the second kind. Due
to this property any state $| \psi \ra$ can be correctly represented
by the analytic function

$$f_{\psi}(z) = \sqrt{I_{2k-1}(2|z|)}/(z^{k-1/2})
\la k,z^{\ast}|\psi\ra,$$
which is of the growth $(1,1)$.  The operators ${K}_{\pm}$ and ${K}_{3}$
act in the Hilbert space of analytic functions $f_{\psi}(z)$ as linear
differential operators

  \be\lb{bgKrep}
{K}_{+} =  z \ , \quad {K}_{-} = 2k \frac{d}{dz} + z
\frac{d^{2}}{dz^{2}} \ , \quad
{K}_{3} = k + z \frac{d}{dz} \ .
    \end{equation}
The $SU(1,1)$ group related CS \ci{KlSk} provide another (analytic in the
unit disk) representation in Hilbert space which has been recently shown
\ci{Vourdas} to be related to the BG representation through a Laplace
transform.

%% Section 3
\section{ The BG CS for $sp(N,C)$ }

The BG CS for semisimple Lie algebras are naturally defined as eigenstates
of mutually commuting Weyl lowering (or raising) operators $E_{\alf^\pr}$
($E^\dg_{\alf^\pr}$) \ci{BaRo}):  $E_{\alf^\pr}|\vec{z}\ra =
z_{\alf^\pr}|\vec{z}\ra$. This can be extended to any algebra, where
lowering/raising operators exist.  We shall consider here the simple Lie
algebra $sp(N,C)$ (the symplectic algebra of rank $N$ and dimension
$N(2N+1)$). We redenote the Cartan-Weyl basis as $E_{ij}, E^\dg_{ij},
H_{ij}$ ($i,j = 1,2,\ldots,N$, $E_{ij}=E_{ji}$, $H_{ij}^\dg = H_{ji}$), and
write the $sp(N,C)$ commutation relations

\be\lb{sp(N,C)comrel}
\bt{l}
$[E_{ij},E_{kl}] = [E^\dg_{ij},E^\dg_{kl}] = 0$,\\
$[E_{ij},E^\dg_{kl}] = \delta_{jk}H_{il} + \delta_{il}H_{jk} +
\delta_{ik}H_{jl} + \delta_{jl}H_{ik}$\\
$[E_{ij},H_{kl}] = \delta_{il}E_{jk} + \delta_{jl}E_{ik}$\\
$[E^\dg_{ij},H_{kl}] = -\delta_{ik}E^\dg_{jl} - \delta_{jk}E^\dg_{il}$\\
$[H_{ij},H_{kl}] = \delta_{il}H_{kj} - \delta_{jk}H_{il}$  \et
\ee \vs{3mm}

The BG CS $|\{z_{kl}\}\ra$ for $sp(N,C)$ are defined as

\be\lb{sp(N,C)bgcsdef}
E_{ij}|\{z_{kl}\}\ra = z_{ij}|\{z_{kl}\}\ra,\quad i,j=1,2,\ldots,N.
\ee
Let us note that the Cartan subalgebra is spanned by $H_{ii}$ only and
$H_{i,j\neq i}$ are also Weyl lowering and raising operators as all
$E_{ij}$ are:  we have simply separated the mutually commuting lowering
operators $E_{ij}$.  We shall construct the $sp(N,C)$ BG CS for the
quadratic bosonic representation, which is realized by means of the
operators

\begin{equation}\label{sp(N,C)bosrep}
E_{ij} = a_ia_j,\quad E^\dg_{ij} = a^\dg_ia^\dg_j,\quad
H_{ij} = \frac{1}{2}(a^\dg_ia_j + a_ja^\dg_i),
\end{equation}
where $a_i,\,a^\dg_i$ are $N$ pairs of boson annihilation and creation
operators. These operators act irreducibly in the subspaces ${\cal
H}^{\pm}$ spanned by the number states $|n_1,\ldots,n_N\ra$ with
even/odd $n_{tot}\equiv n_1+n_2+\ldots+n_N$. The whole Hilbert space
${\cal H}$ of the $N$ mode system is a direct sum of ${\cal H}^{\pm}$.

The $sp(N,C)$ is the complexification of $sp(N,R)$ and
therefor the hermitian quadratures of the above operators span over $R$
the $sp(N,R)$ algebra.  In this way for $N=1$ one gets from
(\ref{sp(N,C)bosrep}) $sp(1,R)\sim su(1,1)$,

\be\lb{su11a}
K_1 = \frac{1}{4}(a^2 + a^{\dg 2}),\,\, K_2 =
\frac{i}{4}(a^2 - a^{\dg 2}),\,\, K_3 = \frac{1}{4}(2a^\dg a +1)
\ee
with the quadratic Casimir operator $C_2 = K_3^2 - K_1^2 - K_2^2 = -3/16$.
Eigenstates of $a^2$ were constructed in the first paper of ref.
\ci{cats1}.

One general property of CS $|\{z_{kl}\}\ra$ for the representation
(\ref{sp(N,C)bosrep}) is that they depend effectively on $N$ complex
parameters $\alf_j$ (not of $N^2+N$ as one might expect).  Indeed, using
the boson commutation relations $[a_i,a_j] =0$ and the definition
(\ref{sp(N,C)bgcsdef}) we easily get

\begin{equation}\label{zijfactoriz}
z_{ij}z_{kl} = z_{ik}z_{jl} = z_{il}z_{jk},
\end{equation}
wherefrom we get the factorization  $z_{ij} = \alf_i\alf_j$. Therefor in
the above bosonic representation the definition (\ref{sp(N,C)bgcsdef}) is
rewritten as

\begin{equation}\label{sp(N,C)bgcsdef1}
a_ia_j|\{\alf_k\alf_l\}\ra = \alf_i\alf_j|\{\alf_k\alf_l\}\ra,
\quad i,j=1,2,\ldots,N.
\end{equation}

The general solution to this system of equations is most easily obtained
in the Glauber CS representation. It reads

\begin{equation}\label{sp(N,C)bgcssolu}
|\{\alf_k\alf_l\}\ra = C_+(\vec{\alf})|\vec{\alf}\ra +
C_-(\vec{\alf})|-\vec{\alf}\ra \equiv |\vec{\alf};C_+,C_-\ra,
\end{equation}
where $|\vec{\alf}\ra$ are multimode Glauber CS, $\vec{\alf} =
(\alf_1,\alf_2,\ldots,\alf_N)$ and $C_\pm(\vec{\alf})$ are
arbitrary functions, subjected to the normalization condition
($|\vec{\alf}|^2 = \vec{\alf}\cdot\vec{\alf}= |\alf_1|^2 + \ldots
+|\alf_N|^2$)

\begin{equation}\label{normcond}
|C_+(\vec{\alf})|^2 + |C_-(\vec{\alf})|^2 +
2{\rm Re}(C_-C^\dg_+)\,N(|\vec{\alf}|) = 1, \quad
N(|\vec{\alf}|)=\la\pm\vec{\alf}|\mp\vec{\alf}\ra = e^{-2|\vec{\alf}|^2}.
\end{equation}
Thus the families of states $|\vec{\alf};C_+,C_-\ra$ represent the whole
set of $sp(N,C)$ BG CS for the representation (\ref{sp(N,C)bosrep}). They
have the form of macroscopic superpositions of Glauber multimode CS. Such
superposition states are also called Schr\"odinger cat states.

The large family of $sp(N,C)$ CS (\ref{sp(N,C)bgcssolu}) contains many
known in quantum optics sets of states \ci{cats1,cats2,Agarwal} and  many
yet not studied sets. Let us  point out some of the well known particular
subsets of (\ref{sp(N,C)bgcssolu}). The limiting cases of $C_-=0$ or
$C_+=0$ recover the overcomplete family of Glauber multimode CS, and
$C_-=\pm C_+$ produces the multimode even/odd CS \ci{cats2}. When $N=2$
the "pair CS" \ci{Agarwal} and the "two mode Schr\"odinger cat states"
\ci{Grobe} are obtained in appropriate manner .  Most of the superpositions
(\ref{sp(N,C)bgcssolu}) for the one mode case ($N=1$) are thoroughly
studied \ci{cats1}. Nevertheless (as far as we know) even in the one
dimensional case no family of Schr\"odinger cat states was pointed out
which is overcomplete in the strong sense in whole Hilbert space ${\cal
H}$.  Here we provide such families.

Consider in (\ref{sp(N,C)bgcssolu}) the choice of
\be\lb{Cpm,phi}
C_+ = \cos\varphi,\,\, C_- = \pm i \sin\varphi,
\ee
which clearly satisfy the normcondition (\ref{normcond}) for any
angle $\varphi$,

\begin{equation}\label{alf;phi1}
|\vec{\alf};\varphi,\pm\ra = \cos\varphi|\vec{\alf}\ra \pm i\sin\varphi
|-\vec{\alf}\ra.
\end{equation}
In Fock basis (number states $|n_1,\ldots,n_N\ra$ we have the expansion

\begin{equation}\label{alf;phi2}
|\vec{\alf};\varphi,\pm\ra = e^{-|\vec{\alf}|^2/2} \sum_{n_i=0}^{\infty}
\frac{\alf_1^{n_1}\ldots\alf_N^{n_N}e^{\pm i
(-1)^{n_1+\ldots+n_N}\varphi}}{\sqrt{n_1!\ldots n_N!}}|n_1,\ldots,n_N\ra.
\end{equation}

By direct calculations we find that these states resolve the unity
operator for any $\varphi$ and thereby provide an analytic representation
in the whole ${\cal H}$,

\begin{equation}\label{alf;phiresolu}
1 = \frac{1}{\pi^N}\int d^2\vec{\alf}
| \vec{\alf};\varphi,\pm\ra \la\pm,\varphi; \vec{\alf}|, \quad
d^2\vec{\alf} = d{\rm Re}\alf_1d{\rm Im}\alf_2 \ldots d{\rm Re}
\alf_Nd{\rm Im}\alf_N .
\end{equation}

States $|\psi\ra$ are represented by  functions

$$f_\psi(\vec{\alf},\varphi,\pm) = e^{|\vec{\alf}|^2/2} \la
\pm,\varphi,\vec{\alf}^{\ast}|\psi\ra, $$
on which the operators $a_j$ and $a^\dg_j$ act as

\begin{equation}\label{aa^dagrep}
a_j = P_\varphi \alpha, \quad a^\dg_j = P_\varphi\frac{d}{d\alpha},
\end{equation}
where $P_\varphi$ acts as inversion operator with respect to $\varphi$:
$P_\varphi f(\varphi) = f(-\varphi)$. At $\varphi = 0, \pi$ the
Glauber CS representation $a=\alf,\, a^\dg = d/d\alf$ is recovered.
The subsets of the one mode states (\ref{sp(N,C)bgcssolu}) corresponding
to $C_- = \pm C_+$ (the even/odd CS $|\alf\ra_{\pm}$ \ci{cats1}) and $C_-
= C_+\exp(i\phi)$ (the Yurke-Stoler states), the nonclassical properties
of which were extensively studied \ci{cats1}, do not resolve the unity in
whole ${\cal H}$. The multimode even/odd CS $|\vec{\alf};C_+,\pm C_+\ra
\equiv |\vec{\alf}\ra_\pm$ \ci{cats2} are overcomplete in the even/odd
subspaces ${\cal H}^\pm$, ${\cal H} = {\cal H}^+ \oplus {\cal H}^-$:

\begin{equation}\label{eocsresolu}
\frac{1}{\pi^N}\int d^2\vec{\alf}|\vec{\alf}\ra_{\pm}\,
_{\pm}\la\vec{\alf}| = 1_{\pm}.
\end{equation}

The notations of (\ref{sp(N,C)bgcsdef}) enable us to introduce the
{\it squared amplitude Schr\"odinger cat states} in the form

\be\lb{sp(N,C)cats1}
|\{z_{kl}\}; D_+,D_-\ra = D_+(\{z_{ij}\}|\{z_{kl}\}\ra +
D_-(\{z_{ij}\}|\{-z_{kl}\}\ra,
\ee
where  the functions $D_{\pm}(\{z_{ij}\}$ have to be subjected to the
normalization condition  (supposing $\la\{z_{kl}\}|\{z_{kl}\}\ra = 1$)
$$
|D_+|^2 + |D_-|^2 + D_-D^*_+\la\{z_{kl}\}|\{-z_{kl}\}\ra +
D^*_-D_+\la\{-z_{kl}\}|\{z_{kl}\}\ra = 1. $$

These superpositions are eigenstates of all $E^2_{ij}$,

\be\lb{sp(N,C)cats2}
E^2_{ij}|\{z_{kl}\};{\rm cats}\ra = z^2_{ij}|\{z_{kl}\};{\rm cats}\ra.
\ee
Evidently, the above two formulas are of quite general nature. But we note
that the above {\it squared amplitude cat's states} are explicitly given,
since $|\{z_{kl}\}\ra$ are constructed (eq. (\ref{sp(N,C)bgcssolu})). In
the two mode case the particular choice of $D_- = D_+\exp(i\phi)$ was
considered in ref. \ci{Grobe} under the name "two mode Schr\"odinger cat
states".

%% Section 4.
\section{BG CS for the algebra $u(p,q)$}

The algebra $u(p,q)$, $p+q=N$, is real form of $sl(N,C)$ and it is
a subalgebra of $sp(N,C)$.  Therefor the BG CS for $u(p,q)$
should be obtained from $sp(N,C)$ CS $|\vec{\alf};C_-,C_+\ra$ by
a suitable restrictions. In this section we consider these
problems in greater detail in the bosonic representation
(\ref{sp(N,C)bosrep}).

The following subset of operators of (\ref{sp(N,C)bosrep}) close
the $u(p,q)$ algebra \ci{BaRo},

\begin{equation}\label{u(p,q)bosrep}
E_{\alf\mu} = a_\alf a_\mu,\quad E^\dg_{\alf\mu} = a^\dg_\alf 
a^\dg_\mu,\quad
H_{\alf\beta} = \frac{1}{2}(a^\dg_\alf a_\beta + a_\beta a^\dg_\alf), \quad
H_{\mu\nu} = \frac{1}{2}(a^\dg_\mu a_\nu + a_\nu a^\dg_\mu),
\end{equation}
where we adopted the notations $\alf,\,\beta,\,\gamma  = 1,\ldots,p$,
$\,\,\, \mu,\,\nu = p+1,\ldots,p+q,\,\,\,\,p+q=N$ (while $i,j,k,l =
1,2,\ldots,N$).  For $p=1 =q$ the three standard $su(1,1)$ operators
$K_\pm,\,K_3$ are $K_- = E_{12}$, $K_+ = E^\dg_{12}$, $K_3 = (H_{11}
+H_{22} + 1)/2$.  The subsets of hermitian operators

\be\lb{u(p),u(q)bosrep}
\bt{l}
$M^{(p)}_{\alf\bet} =\frac{1}{2}(H_{\alf\bet}+H_{\bet\alf}-
\dlt_{\alf\bet}),
\,\, \tld{M}^{(p)}_{\alf\bet} = i(H_{\bet\alf}- H_{\alf\bet})$,\\
$M^{(q)}_{\mu\nu} =\frac{1}{2}(H_{\mu\nu}+H_{\nu\mu}-\dlt_{\mu\nu}),
\,\, \tld{M}^{(q)}_{\mu\nu} = i(H_{\nu\mu}- H_{\mu\nu})$ \et
\ee
realize representations of compact subalgebras $u(p)$ and $u(q)$
correspondingly. The $u(p,q)$ (\ref{u(p,q)bosrep})  algebra acts
irreducibly in the subspaces of eigenstates of the hermitian operator $L$,

\be\lb{L}
L = \sum_\alf M^{(p)}_{\alf\alf} -
\sum_\mu M^{(q)}_{\mu\mu} = \sum _\alf H_{\alf\alf} - \sum_\mu
H_{\mu\mu} .  \ee

This is the linear in generators Casimir operator and the higher Casimirs
here are expressed in terms of $L$ \ci{ITT}).  Denoting the eigenvalue of
$L$ by $l$ we have the expansion ${\cal H} = \sum_{l=-\infty}^{\infty}
\oplus{\cal H}_l$. The representations corresponding to $\pm l$ are
equivalent (but the subspaces ${\cal H}_{\pm l}$ are orthogonal).

The commuting Weyl lowering generators of $u(p,q)$ are $E_{\mu\alf} =
a_\mu a_\alf$. Therefor the $u(p,q)$ BG CS are defined (in the above
bosonic representation) as $|\{\alf_\bet\alf_{\nu}\};l,p,q\ra$,

\be\lb{u(p,q)bgcsdef}
\bt{l}$a_\mu a_\gamma |\{\alf_\bet\alf_nu\};l,p,q\ra = \alpha_\mu
\alpha_\gamma |\{\alf_\bet\alf_nu\};l,p,q\ra,$\\
                                        \\[-2mm]
$\gamma=1,\ldots,p,\,\,\,\,\mu=p+1,\ldots,p+q.$ \et \ee

Solutions to these equations are ($|\!|\vec{\alf};l,p,q\ra =
|\!|\{\alf_\bet\alf_nu\};l,p,q\ra$)

\begin{equation}\label{u(p,q)bgcssolu}
|\!|\vec{\alf};l,p,q\ra =
\sum_{\tld{n}_p-\tld{n}_q=l}
\frac{\alf_1^{n_1}...\alf_N^{\tld{n}_p-\tld{n}^\pr_q-l}}{\sqrt{n_1!...
n_{N-1}!(\tld{n}_p-\tld{n}^\pr_q-l)!}} |n_1,...,n_{N-1}; \tld{n}_p-
\tld{n}^\pr_q-l\ra,
\end{equation}
where $\alpha_i$, $i=1,\ldots,N$, are arbitrary complex parameters,
$\tld{n}_p = \sum_\alf n_{\alf}$, $\tld{n}_q = \sum_\mu n_\mu$,
$\tld{n}^\pr_q = n_1 + n_2 + \ldots + n_{N-1}$ and $l = \tld{n}_p
- \tld{n}_q$ ($|\!|\psi\ra$ denotes a nonnormalized state). If we multiply
$|\!|\vec{\alf};l,p,q\ra$ by $\exp(-|\vec{\alpha}|^2/2)$ and sum over $l$
we evidently get the normalized Glauber multimode CS $|\vec{\alpha}\ra$
(for any pair $p,\,q$),

\begin{equation}\label{ccsexpan}
|\vec{\alf}\ra =
e^{-\frac{1}{2}|\vec{\alf}|^2}\sum_{l=-\infty}^{\infty}
|\!|\vec{\alf};l,p,q\ra .  \end{equation}

The last equality suggests that the states
$|\!|\vec{\alf};l,p,q\ra $ form overcomplete families in ${\cal H}_l$ for
every $p,\,q$. And this is the case. Indeed, let $|\psi_l\ra$ be any
states from ${\cal H}_l$. Using the overcompleteness of $|\vec{\alf}\ra$,
the formula (\ref{ccsexpan}) and the orthogonality relations

\begin{equation}\label{orthogo1}
\la p,q,l^\pr;\vec{\alf}|\!|\vec{\alf};l,p,q\ra = \delta_{l^\pr l},
\end{equation}

one can get the resolution of unity in ${\cal H}_l$ in terms of the
$u(p,q)$ CS  $|\!|\vec{\alf};l,p,q\ra$,

\begin{equation}\label{u(p,q)resolu1}
  \int d\mu(\vec{\alf})
|\!|\vec{\alf};l,p,q\ra\la p,q,l;\vec{\alf}|\!| = 1_{l}, \quad
d\mu(\vec{\alf}) = \frac{1}{\pi^N} d^2\vec{\alf}e^{-|\vec{\alf}|^2} .
\end{equation}

Now we note that in $u(p,q)$ CS (\ref{u(p,q)bgcssolu}) one complex
parameter, say $\alpha_N$, can be absorbed into the normalization factor by
redefining the rest as $z_1 = \alpha_1\alpha_N, \ldots, z_p =
\alpha_p\alpha_N$, $z_{p+1} = \alpha_{p+1}/\alpha_N, \dots z_{N-1} =
\alpha_{N-1}/\alpha_N$. Then we can write $|\!|\vec{\alf};l,p,q\ra\ =
|\!|\vec{z};l,p,q\ra$,

\begin{equation}\label{z-u(p,q)bgcs}
|\!|\vec{z};l,p,q\ra = \sum_{\tld{n}_p-\tld{n}_q=l}
\alf_N^{-l}\frac{z_1^{n_1}...z_{N-1}^{N-1}}{\sqrt{n_1!...
n_{N-1}!(\tld{n}_p-\tld{n}^\pr_q-l)!}}
|n_1,...,n_{N-1},\tld{n}_p-\tld{n}^\pr_q-l\ra,
\end{equation}
where $\vec{z} = (z_1, \ldots, z_{N-1})$.
In terms of variables $\vec{z}$ the resolution of unity reads
($d^2\vec{z}= d{\rm Re}z_1d{\rm Im}z_1 \ldots d{\rm Re}z_{N-1}d{\rm
Im}z_{N-1}$)

\begin{equation}\label{u(p,q)resolu2}
\bt{l}$1_l = \int d\mu(\vec{z};l,p,q)|\!|\vec{z};l,p,q\ra\la
q,p,l;\vec{z}|\!|,$\\[-2mm]
                     \\
$d\mu(\vec{z},l,p,q) =
\frac{1}{\pi^{N}}F(|\vec{\tld{z}_p}|,|\vec{\tld{z}_q}|;l,p,q)d^2\vec{z}$,
\et \end{equation}

where $|\vec{\tld{z}_p}|^2 = |z_1|^2 + \ldots + |z_p|^2$,
$|\vec{\tld{z}_q}|^2 = |z_{p+1}|^2 + \ldots + |z_{N-1}|^2$, and

\begin{equation}\label{F}
F(|\vec{\tld{z}_p}|,|\vec{\tld{z}_q}|;l,p,q) = \int
d^2\alpha_N|\alpha_N|^{2(q-p-l)}\exp\l[-\l(\frac{|\vec{\tld{z}_p}|^2}
{|\alf_N|^2} + |\vec{\tld{z}_q}|^2|\vec{\alf}_N|^2 + |\alf_N|^2\r)\r]
\end{equation}
One can prove that the above measure is unique in the class of continuous
functions (see the Appendix). In the particular
case of $q=1$ (then $p = N-1$, $\tld{n}^\pr_q = 0$ and $\vec{z}_q = 0$)
and negative $l$, $-l\geq p$, the $u(p,1)$ BG CS were constructed in
\ci{Fujii} with the resolution unity measure (in ${\cal H}_l$)

\be\lb{}
d\mu^\pr(\vec{z}) = F^\pr(|\vec{z}|,l,p,1)d^2\vec{z},\quad
F^\pr = \frac{2|\vec{z}|^{-l-p}}{\pi^N} K_{-l-p}(2|\vec{z}|),
\ee
where $K_\nu(z)$ is the modified Bessel function of the second kind.
From continuity of $F^\pr(|\vec{z}|,l,p,1)$ and $F(|\vec{z}|,l,p,1)$ we
deduce that they coincide (see the Appendix). Then using the analyticity
property of Bessel functions $K_\nu(z)$ \ci{Stegun} we establish the
following integral
%%On positive half line $z^\nu K_\nu(z)$ is continuous for any
%%$\nu$ (and positive for $\nu>-1$) \ci{Stegun}. Our function
%%$F(|\vec{z}_p|,l,p,1)$ is continuous (and positive) for any $l$ as one
%%can easily see.
representation for $K_\nu(z)$ with $\nu = 0,1,\ldots$ and ${\rm
Re}z \geq 0$,

\be\lb{K_nu}
 K_\nu(2z) = 2\pi(2z)^{-\nu}\int_{0}^{\infty}dx\,x^{1+\nu}e^{-(x + z^2/x)}.
\ee
For $p=1,\,q=1$ our states $|\vec{z};l,p,q\ra$ recover (as the states of
\ci{Fujii} do) the BG CS for $u(1,1)$ \ci{BG}.

%% Section 5
\section{Discussion}

We have constructed the Barut-Girardello type coherent states (BG CS) for
the noncompact algebras $u(p,q)$ and $sp(N,C$ in the $N$ mode quadratic
bosonic representation (Eq. (\ref{sp(N,C)bosrep})). The general set of
$sp(N,C)$ CS is obtained in the form of macroscopic superpositions
$|\vec{\alf};C_+,C_-\ra$ (Eq.  (\ref{sp(N,C)bgcssolu})) of multimode
Glauber CS. Such superposititions are called ordinary multimode
Schr\"odinger cat states.  Several particular cases of these
cat states are intensively studied in the literature, due to their
nonclassical properties \ci{cats1,cats2}.  The new states (e.g.
$|\vec{\alf};\varphi,\pm\ra$, eq. (\ref{alf;phi1})) can also exhibit
interesting nonclassical properties, such as ordinary quadrature squeezing
and subpoissonian photon statistics.  They possess the intelligence
property to minimize the Robertson multidimensional uncertainty relation
\ci{Rob} for the hermitian quadratures $X_{ij},\,Y_{ij}$ of Weyl lowering
operators $E_{ij}=X_{ij}-iY_{ij}$, since they are eigenstates of all
$E_{ij}$ \ci{Trif97}. The constructed $u(p,q)$ and $sp(N,C)$ BG CS are
stable under the action of the free field evolution operator.

However, these states cannot exhibit squeezing of the observables $X_{ij}$
and $Y_{ij}$ since here the variances of $X_{ij}$ and $Y_{ij}$ are equal
as a result of their eigenvalue property \ci{Trif94}.  In the
representation $E_{ij} = a_ia_j$ the $X_{ij}$ (or $Y_{ij}$) squeezing is
the {\it squared amplitude squeezing}.

Squeezing of the quadratures of
$a_ia_j$ for a given modes $i,\,j$ can be achieved in two ways:\\
a) in the eigenstates of the complex combinations $u_{ij}a_ia_j +
v_{ij}a^\dg_ia^\dg_j$;\\
b) in the eigenstates $a^2_ia^2_j$.

For $i\neq j$ squeezed states (SS) of type a) were constructed in
\ci{Trif94} as eigenstates of $uK_- + vK_+$, $K_\pm$ being generators of
$SU(1,1)\sim Sp(1,R)$ in the representations with Bargman index
$k=1/2,1,\ldots$. Those eigenstates minimize the Schr"odinger
uncertainty relation for $K_1$ and $K_2$. By means of two boson operators
$a_i,\,a_j$ the realization of $K_\pm$ is according to
(\ref{sp(N,C)bosrep}) with fixed $i,\,j$ and $k =
(1-l)/2,\,\,\,l=n_i-n_j$. Eigenstates of other combinations of $K_\pm$ and
$K_3$ in the two mode case were studied in \ci{Grobe,BrifBen,Perina}.  In
the one mode case Schr\"odinger intelligent states for two generators of
$SU(1,1)$ were constructed and discussed in \ci{Trif95,Brif96}.  As it was
noted in \ci{Trif97} the passage from eigenstates of $E_{ij}$ (i.e.  from
BG CS) to the eigenstates of combination  $u_{ij}a_ia_j +
v_{ij}a^\dg_ia^\dg_j$ (which could exhibit squeezing of $X_{ij},\,Y_{ij}$)
can't be performed by unitary squeeze operator $S(u,v)$. Let us recall
that the ordinary SS are associated to the complexified
Heisenberg algebra, where the passage from eigenstates of $a_i$ to the
eigenstates of $u_{j}a_j + v_ja^\dg_j$ is accomplished by unitary operator
- the standard squeeze operator.

 A family of states in which squeezing of quadratures of any product
$a_ia_j$, $i,j = 1,2,\ldots,N$, can occur should be called a family of
{\cal multimode squared amplitude SS}. Example of such multimode SS is
given by the Robertson intelligent states, which should be eigenstates of
$u_{kl;ij}a_ia_j + v_{kl;ij}a^\dg_ia^\dg_j$ (summation over repeated
indices) \ci{Trif97}. These are multimode SS of type a). Multimode squared
amplitude SS of type b) are eigenstates of all $(a_ia_j)^2$. The latter
have been introduced here explicitly by means of eq.
(\ref{sp(N,C)cats1}).  For $N=2$ a particular subsets of
(\ref{sp(N,C)cats1}) have been studied in \ci{Grobe}, where it was shown
that they exhibit ordinary squeezing and subpoissonian statistics. We note
here that they can exhibit squared amplitude squeezing as well, which
should be considered elsewhere.

As to the overcompleteness properties of the constructed CS it is worth
noting that the families of $sp(N,C)$ CS $|\vec{\alf};C_+ =
\cos\varphi,C_- = \pm i\sin\varphi\ra \equiv |\vec{\alf};\varphi,\pm\ra$
are {\it overcomplete in the whole Hilbert space}, while the
representation (\ref{sp(N,C)bosrep}) is irreducible in the subspaces
${\cal H}^\pm$ (spanned by Fock states with even/odd total number of
bosons) only. Let us compare this property with the corresponding one of
group related CS - by construction the latter are overcomplete in the
space which is irreducible under the group (and therefor algebra) action:
the $Sp(N,R)$ group related CS in the representation (\ref{sp(N,C)bosrep})
are overcomplete in ${\cal H}^\pm$, not in the whole ${\cal H}$. The BG
type of CS are in fact algebra related and enable the resolution of unity
in the whole ${\cal H}$.  The resolution unity measure for the $sp(N,C)$
normalized BG CS $|\vec{\alf};\varphi\pm\ra$ was obtained the same as for
the multimode Glauber CS. The resolution unity measures for $u(p,q)$ CS
$|\!|\vec{z};l,p,q\ra$ generalize the BG CS measure \ci{BG} for $u(1,1)$
and the recently obtained measure \ci{Fujii} for $u(p,1)$ to the case of
any $p,q$ and any value of the first Casimir $l$.  Finally it worth noting
high symmetry of Glauber multimode CS: these are simultaneously $H_W$
group related CS and $h_W$ and $sp(N,C)$ algebra related CS (where $H_W$
denotes the Heisenberg-Weyl group).

\section{Appendix}
\bc {\bf A. Uniqueness of the resolution unity measures
$d\mu(\vec{z},l,p,q)$} \ec

Suppose that there exists another function $F^\pr(|\vec{\tld{z}_p}|,
|\vec{\tld{z}_q}|;l,p,q)$ such that the new measure $d\mu^\pr = F^\pr
d^2\vec{z}$ resolves the unity $1_l$ as in eq.  (\ref{u(p,q)resolu2}).
Then we should have

\be\lb{}
0 = \int d^2\vec{z}\l
[F(|\vec{\tld{z}_p}|,|\vec{\tld{z}_q}|;l,p,q)-
F^\pr(|\vec{\tld{z}_p}|,|\vec{\tld{z}_q}|;l,p,q)\r]
|\!|\vec{z};l,p,q\ra\la q,p,l;\vec{z}|\!|.  \ee

Substituting the expansion (\ref{}) of $|\!|\vec{z};l,p,q\ra\la$, and
integrating with respect to angles $\varphi_i=\arg z_i$ we obtain that the
difference function

$$\Phi(r_1,r_2,\ldots,r_{N-1})\equiv
F(\tld{r}_p,\tld{r}_q;l,p,q)- F^\pr(\tld{r}_p,\tld{r}_q;l,p,q),$$

where
$\tld{r}_p\equiv |\vec{\tld{z}_p}|=\sqrt{r_1^2+...+r_p}$ and
$\tld{r}_q\equiv|\vec{\tld{z}_q}|=\sqrt{r^2_{p+1}+...+r^2_{N-1}}$, should
be orthogonal to the monomials

$$r_1^{2n_{1}+1}\ldots r_{N-1}^{2n_{N-1}+1},\,\,\,
r_i=|z_i|,\,\, i=1,\ldots,N-1,\,\, n_i=1,\ldots.$$
Redenoting $r_i^2$ again as $r_i$ one can write this orthogonality in the
form

\be\lb{orthogo2}
\int_{0}^{\infty} dr_1\ldots dr_{N-1}\Phi(r_1,\ldots,r_{N-1})
r_1^{n_1}\ldots r_{N-1}^{n_{N-1}} = 0,
 \ee
where $ n_i = 1,2,\ldots,\quad i=1,\ldots,N-1$.
Herefrom  it follows that $\Phi$ is orthogonal to any function $f(r_1\ldots
r_{N-1})$ which admits Teylor expansion,

 \be\lb{orthogo3}
\int_{0}^{\infty} dr_1\ldots
dr_{N-1}\Phi(r_1,\ldots,r_{N-1}) f(r_1\ldots r_{N-1}) = 0. \ee

This implies that $\Phi = F-F^\pr = 0$ almost everywhere. Indeed, if
$\Phi\neq 0$ it must be nonpositive definite (in order to obey
(\ref{orthogo2})) and if $\Phi$ is well behaved (it is sufficient to be
continuous) we could find $f$ which is negative in the domains where
$\Phi<0$.  But then  we could not maintain (\ref{orthogo3}), which proves
that $F =F^\pr$ almost everywhere.  If $F$ and $F^\pr$ are continuous they
should coincide.  \vs{1cm}

\bc {\bf B. Proof of the representation (\ref{K_nu}) of $K_\nu(z)$} \ec

In case of $q=1$ ($p=N-1$) and $-l\geq p$ our measure function $F$, eq.
(\ref{F}),  depends on $r_1,\ldots,r_p$ through $|\vec{z}| =
[|z_1|^2+\ldots+|z_{p}|^2]^{-1/2} \equiv \tld{r}_p$ and it is easily seen
that $F$ is a continuous (and positive) function of $r_1,\ldots,r_p$.  The
measure function of ref.  \ci{Fujii} is $F^\pr \sim
r_p^{-l-p}K_{-l-p}(2r_p)$ (in their case $-l>p$) is also continuous
\ci{Stegun}, therefor the difference $\Phi$ of these two functions is
continuous and in view of (\ref{orthogo2}) they have to coincide
pointwise. This proves formula (\ref{K_nu}) for
${\rm Im}z =0,\,\,{\rm Re}z > 0$.

The Bessel function $K_\nu(z)$ is analytic and regular everywhere except
of the negative half of the real line \ci{Stegun}. Let us consider the
right hand side of (\ref{K_nu}) as a definition of a new function
$F(z;\nu)$, $z$ complex, $\nu$ real. The integral is convergent for Re$z >
0$ and the function $F(z;\nu)$ is evidently analytic. We proved in the
above that the two analytic functions $F(z;\nu$ and $K_\nu(2z)$ ($\nu =
0,1,\ldots$) coincide on the positive part of the real line.  Then they
coincide in the whole domain of analyticity.
We note that (\ref{K_nu}) does not hold for negative $\nu$:  $K_{-\nu}(2z)
= K_{\nu}(2z)$, but $F(z;-\nu)\neq F(z;\nu)$.
%%%%

\end{document}